\documentclass[aip,reprint,floatfix,twocolumn]{revtex4-2}

\usepackage{graphicx}
\usepackage{physics}
\usepackage{bm}
\usepackage{lipsum}
\usepackage{color}
\usepackage[utf8]{inputenc}
\usepackage[T1]{fontenc}
\usepackage{mathptmx}
\usepackage{etoolbox}

\begin{document}

\title{\textcolor{black}{Experimental Plasma Density Profiles Determined Through Measurements of the Magnetosonic Wave  Speed}} %

\author{C. Kuchta}
\email[]{ckuchta@wisc.edu}
\author{J. Egedal}%
\author{A. Mhatre}
\author{P. Gradney}
\author{J. Olson}
\author{X. Yu}
\author{C. Forest}
\affiliation{ 
Department of Physics, University of Wisconsin-Madison, Madison, Wisconsin 53706, USA
}

\date{9 September 2026}

\begin{abstract}
Information on plasma density in laboratory plasmas is commonly acquired using either Langmuir probes or optical diagnostics. Here, we present an alternative approach, inferring the density profile from magnetic measurements of a plasma wave. In particular, during the process of creating a reconnecting current layer for magnetic reconnection experiments in the Big Red Ball, the reconnection drive first launches a large amplitude fast magnetosonic wave. The propagation of the wavefront is measured with high spatial and temporal resolutions by {\sl in situ} magnetic diagnostics. Given a known uniform background magnetic field strength and the known dispersion relation of the wave, we here show how the characteristics of the wavefront can be applied to determine the initial plasma density profile.
\end{abstract}

\maketitle %

\section{Introduction}

In laboratory plasma experiments, the profile of the plasma number density is often key to understanding and interpreting the plasma dynamics. Depending on the plasma size and plasma parameters, the plasma density can typically be inferred through optical diagnostics or through the application of {\sl in situ} diagnostics such as Langmuir probes. In particular, for basic plasma physics experiments, Langmuir probes are commonly used given their simplicity of implementation. Nevertheless, the Langmuir probe measurements require data at multiple probe biases in order for the plasma density to be determined, adding some limitations to their use.
Furthermore, these probes often only take measurements at a single spatial location, and it can also be challenging to accurately infer the desired plasma parameters from the probe measurements.
On the other hand, B-dot coils have a simple conversion from raw measurement to magnetic field values and are used in many devices\cite{gekelman_2016}\cite{olson_2021}. Some machines, such as the Big Red Ball (BRB)\textcolor{black}{, a 3 m diameter ring cusp confinement plasma device}, use arrays of B-dot coils to measure the change in field for each discharge in many spatial locations. 

The Terrestrial Reconnection EXperiment (TREX) is a particular configuration implemented in the Big Red Ball (BRB) within the Wisconsin Plasma Physics Laboratory (WiPPL). The configuration is applied to the study of fast magnetic reconnection in a collisionless setting relevant to reconnection in the Earth's magnetosphere\cite{olson_2021}\cite{gradney_2023}. 
The TREX configuration is compatible with physical diagnostics, and given the relatively large plasma volume, the device is well suited for the application of both B-dot and Langmuir probes. 

For driving reconnection, TREX applies a strong reconnection drive which includes the sudden energization of a system of coils encircling the plasma. Before the formation of the reconnection current layer, this reconnection drive produces a strong magnetosonic wave including a wavefront moving radially inward towards the center of the device\cite{olson_2020}\cite{gradney_2023}. The propagation speed of this front is governed by the dispersion relation of the magnetosonic wave, directly related to the plasma density. Below we will discuss and show how the analysis of the wave propagation provides reliable density profiles which are consistent with measurements by Langmuir probes. These profiles characterize the initial plasma produced by an array of plasma guns, which is valuable for analyzing and understanding the plasma dynamics at later times during the plasma discharges.

\begin{figure}
    \includegraphics[width=\linewidth]{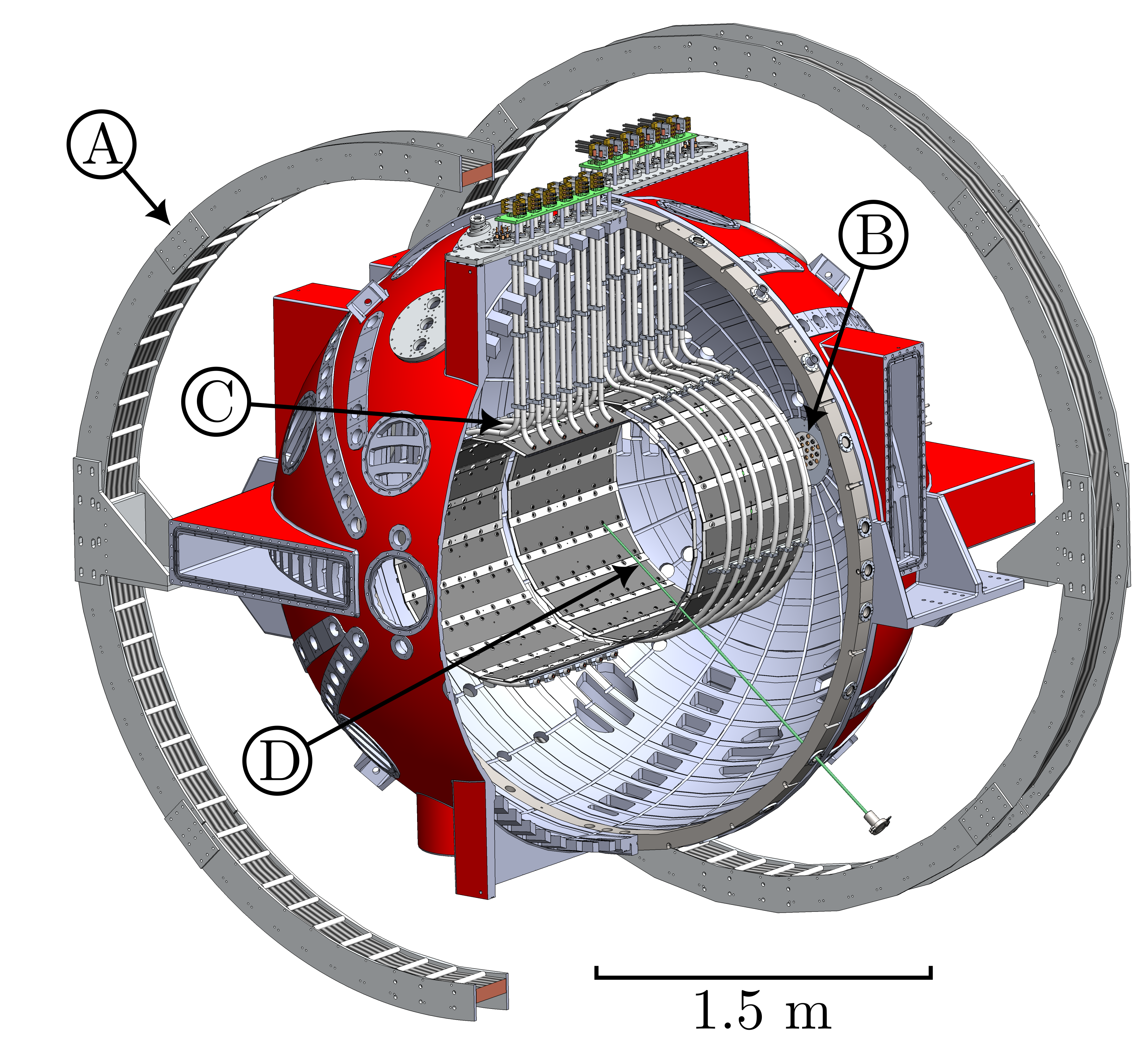}
    \caption{A CAD diagram of the BRB. Each discharge starts with a background axial field (up to 15 mT) created by the Helmholtz coil (A). Plasma is injected from the pole of the machine (up to $10^{19} \, \text{m}^{-3}$ and $4$ eV) (B). Reconnection is driven by applying a large voltage across the drive cylinder (C). Plasma dynamics are measured by either Langmuir or B-dot probes (D).}
    \label{fig:BRB-CAD}
\end{figure}

\section{Experimental Overview} \label{sec:experimental_overview}

\begin{figure}
    \includegraphics{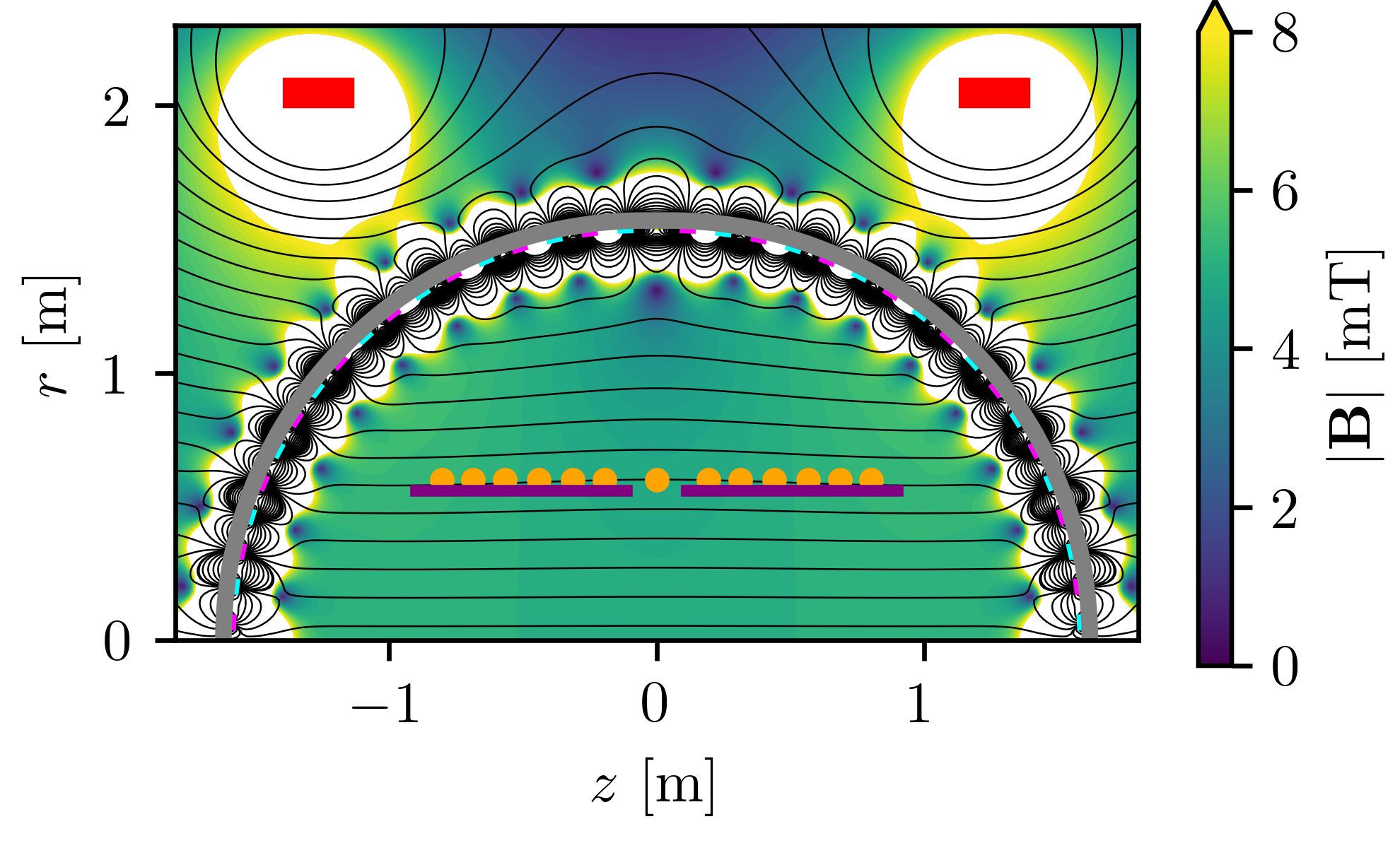}
    \caption{The theoretical initial field profile from the permanent magnets (purple and cyan rectangles) and Helmholtz coil (red rectangle). The location of the drive cylinder and its coils are shown in purple and orange respectively. The plasma guns, located at $r = 0$, $z = 1.5$ m inject plasma that balances the initial field.} %
    \label{fig:initial-fields}
\end{figure}

For the present experiments the plasma in the BRB is produced by an array of plasma guns\cite{Fiksel_Almagri_Craig_Iida_Prager_Sarff_1996} mounted at one of the poles of the spherical 3 m diameter vacuum vessel. To enhance the plasma confinement, rings of permanent magnets at the wall create a multicusp configuration. This cusp field decreases rapidly with distance and has negligible impact on the dynamics closer to the center of the device dominated by a uniform background field along the $z$-axis produced by a Helmholtz coil, Fig.~\ref{fig:BRB-CAD}(A)\cite{forest_2015}.
Additionally, we may apply a $\phi$ directed field, produced by a coil running along the $r = 0$ center line of the BRB.
We then inject plasma from plasma guns located at the $+z$ pole of the machine, Fig.~\ref{fig:BRB-CAD}(B), creating a screw-pinch equilibrium.
\textcolor{black}{The detailed magnetic field geometry of the permanent magnets at the wall, as well as the Helmholtz coil, is shown in Fig.~\ref{fig:initial-fields}}. 
    \textcolor{black}{In vacuum, the drive cylinder, shown schematically in Fig.~\ref{fig:BRB-CAD}(C) and indicated by the orange circles in Fig.~\ref{fig:initial-fields}, creates a uniform magnetic field anti-parallel to the Helmholtz field\cite{gradney_2023}}. 

\begin{figure*}
    \centering
    \includegraphics{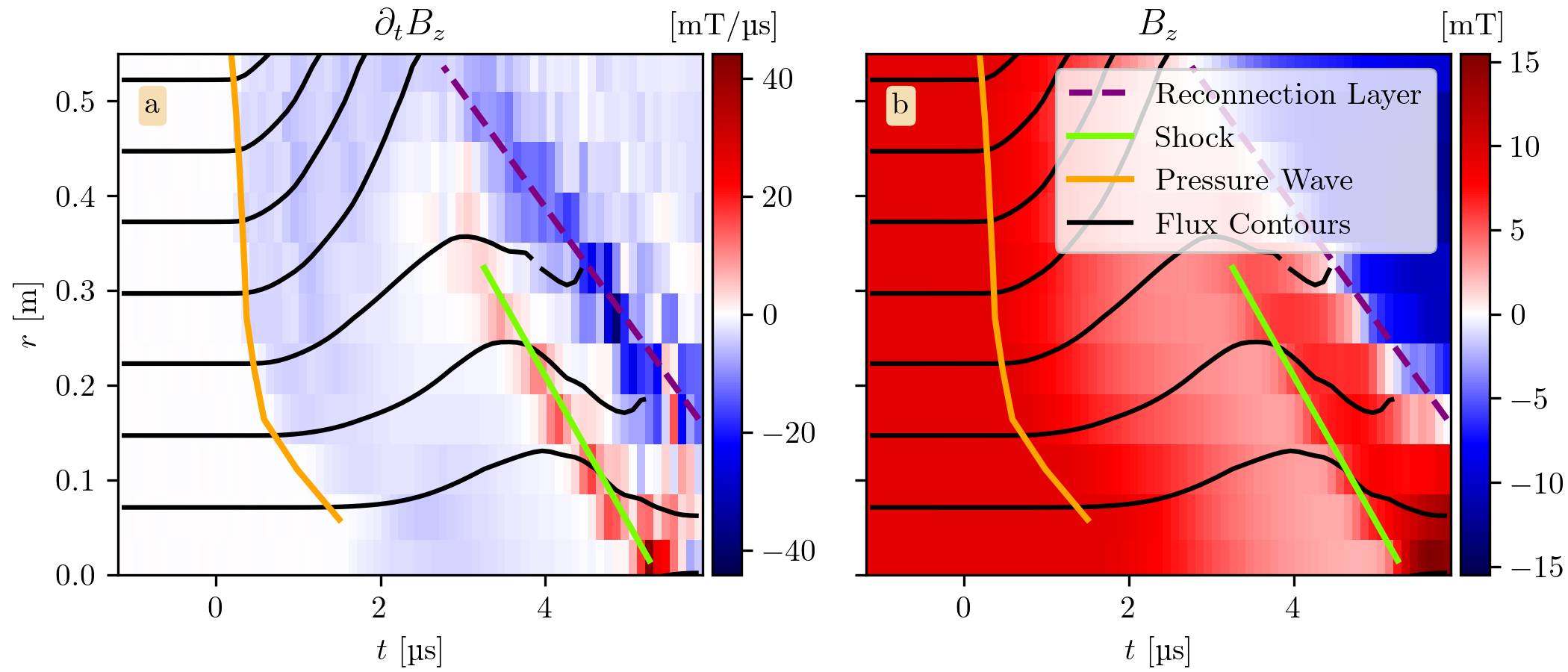}
    \caption{Experimental measurements from a single shot where the probes are at $z=0$. The drive cylinder at $r = 0.6$ m turns on at $t = 0$. This releases the magnetic pressure at high radii allowing the plasma to expand which leads to a pressure wavefront moving radially inwards (orange). \textcolor{black}{This wavefront is used for inferring the initial density.} \textcolor{black}{Afterwards}, a shock interface (green) is formed in front of the reconnection layer (purple dashed) which moves radially inwards\textcolor{black}{\cite{olson_2021}}. \textcolor{black}{The straight lines are linear fits to the location where $\partial_t B_z$ is maximized and $B_z = 0$ respectively.} Once the wavefront passes, the magnetic field lines move outwards following the flux contours (black).}
    \label{fig:experimental-evolution}
\end{figure*}

\textcolor{black}{
In our reconnection experiments, the drive cylinder is filled with plasma prior to the energization of the cylinder. This yields an evolution of the plasma and magnetic fields which are much more involved. In the present work, we will focus on two diagnostics: a linear array of B-dot sensors extending along the radius of the drive cylinder, Fig.~\ref{fig:BRB-CAD}(D), and a multi-tip Langmuir probe that can be moved to any position in the device\cite{olson_2020}.
An example of the magnetic data that forms the basis for the analysis is shown in
Fig.~\ref{fig:experimental-evolution}. The color map in panel (a) shows $\partial B_z/ \partial t$ recorded as a function of $t$ and $r$ measured by the Bdot probes at $z=0$. Characteristic features in the data are identified by three lines. The features marked by lines first visible around $t=3$ \textmu s, correspond to a shock layer (green) moving radially inwards ahead of the reconnection layer (maroon). These features are discussed elsewhere \cite{olson_2021,Sam_Greess_2021} and include highly nonlinear plasma dynamics not described by the present analysis.} 

\textcolor{black}{The present analysis is only concerned with the initial wave dynamics identified in Fig.~\ref{fig:experimental-evolution}(a) by the orange line,   This feature corresponds to the disturbance launched by the  drive cylinder at $r\simeq 0.6$ m being energized at $t=0$ \textmu s, and is readily identified in the raw $\partial B_z/ \partial t$ prior to both shock formation and reconnection.
As shown in Fig.~\ref{fig:experimental-evolution}(b), at the time and location of the orange line the $B_z$ magnetic field is still undisturbed with a value corresponding to the applied background Helmholtz field.}

The speed, $v_{\textcolor{black}{\text{meas}}}$, 
\textcolor{black} { of the initial plasma disturbance is readily found by dividing the distance between probes by the time delay for when the feature in $\partial B_z/\partial t$ first appears for each probe (this will be discussed in more details below). As will also be shown later, this propagation through the unperturbed plasma is approximately the local Alfv\'{e}n speed which results in a calculated number density, $n_0$, of}
\begin{equation}
\label{eq:n0}
    \color{black} n_0 = \frac{B_H^2}{\mu_0} \frac{1}{m_i v_{\text{meas}}^2}
\end{equation}
\textcolor{black}{where $B_H$ is the Helmholtz field and $m_i$ is the ion mass.}
\textcolor{black}{
Justifications for using this Alfv\'{e}n speed approximation is discussed in Sections.~\ref{sec:wave_simulation} and \ref{sec:data_evaluation}.} 

Since the plasma is injected along the axis of the device, the plasma must diffuse across field lines to reach a high radius. Consequently, the density often becomes too low to be accurately characterized by the electrostatic probes. Additionally, the applied Langmuir probe can only measure the density at a single location for each shot, which makes it time consuming to characterize the density at many different radii. We also lack a method for absolute calibration of these probes, providing additional motivation for developing the new technique to obtain an absolute measurement of the plasma density profile.

In the following Sections, we will discuss our detailed approach for calculating the initial plasma density using magnetic field data owing to a magnetosonic wave that travels radially inward. Section \ref{sec:wave_simulation} describes a linearized magnetohydrodynamic (MHD) simulation of our system when the drive cylinder is energized \textcolor{black}{after which we describe the calculation of the density from the B-dot measurements (Section \ref{sec:data_evaluation}}. We then compare our simulation results to the experimental data (Section \ref{sec:procedure_validation}) and discuss their implications (Section \ref{sec:conclusions}).

\section{Wave Simulation}\label{sec:wave_simulation}

\begin{figure*}
    \centering
    \includegraphics{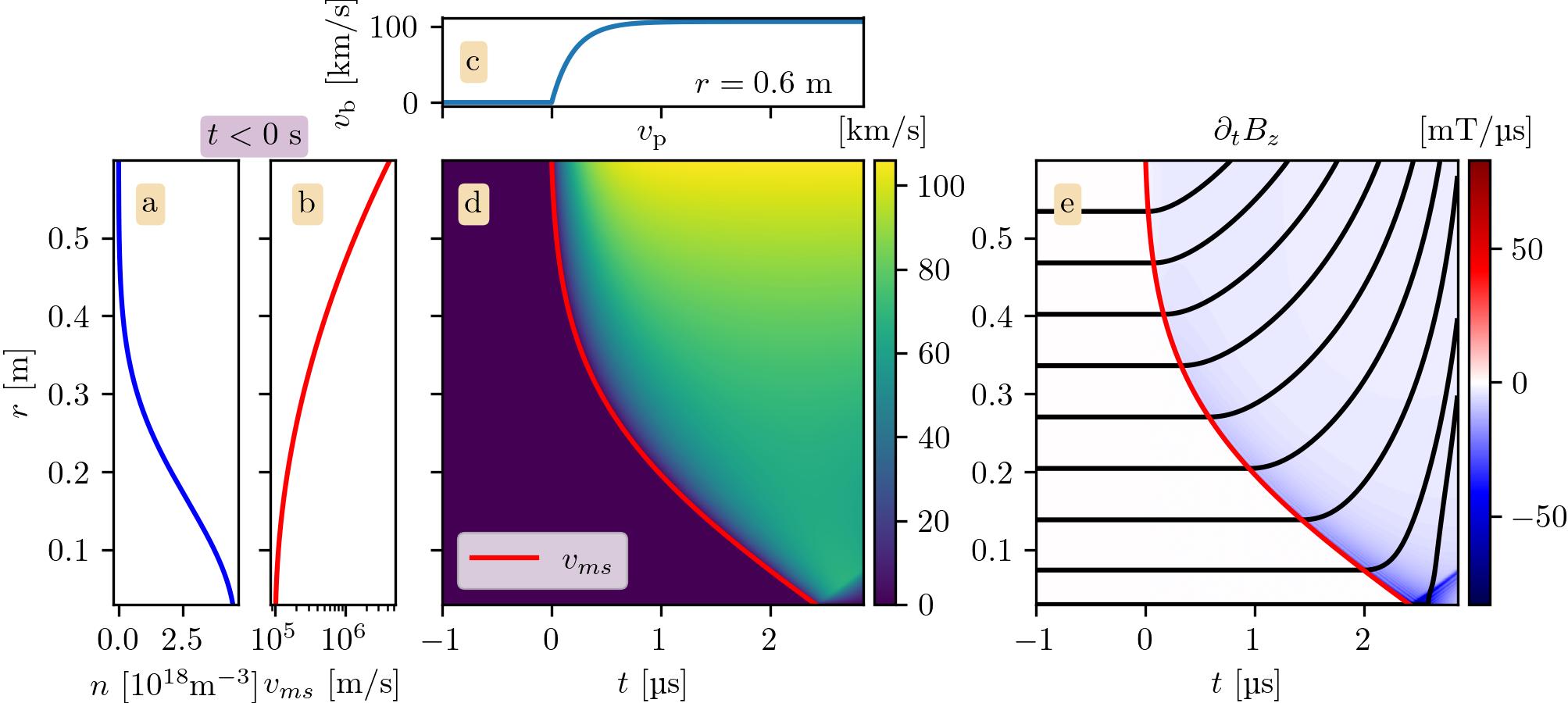}
    \caption{The simulated wavefront from a 1D linearized MHD simulation. Panel a) shows the initial density profile and panel b) shows the fast magnetosonic speed (red). Panel c) is the boundary velocity set by the applied loop voltage. Panel d) is the velocity of the plasma which only begins to move outward once the wavefront passes with the magnetosonic streamline drawn in red \textcolor{black}{with initial and boundary conditions set by the density a) and boundary velocity c)}. Panel e) shows the \textcolor{black}{simulated} measurements from a B-dot coil and flux contours which are similar to \ref{fig:experimental-evolution}, a).}
    \label{fig:simulation-setup}
\end{figure*}

To understand the properties of how the initial magnetic perturbation propagates inwards from the drive cylinder, we start by linearizing the MHD equations as in Parker\cite{Parker_2020}(appendix A). However, our derivation departs at equation (A13) leaving us with the following:
\begin{equation}
\begin{aligned}
\label{eq:vector_linearized_MHD}
    \rho_0 \partial_{tt} \bm{\xi} =& \frac{1}{\mu_0}\left\{\nabla \times \left[\nabla \times \left(\bm{\xi} \times \vb{B}_0\right)\right]\right\}\times \vb{B}_0 + \frac{1}{\mu_0} \left(\nabla \times \vb{B}_0\right) \\
    &\times \left[\nabla \times \left(\bm{\xi} \times \vb{B}_0\right)\right] + \nabla \left(\bm{\xi} \cdot \nabla p_0 + \gamma p_0 \nabla \cdot \bm{\xi}\right)
\end{aligned}
\end{equation}
where we have written the differential equation in terms of $\bm{\xi}$, the plasma displacement vector ($\bm{v} = \partial_t \bm{\xi}$). $\rho_0$ is the mass density, $\vb{B}_0$ is the initial magnetic field, $p_0$ is the plasma pressure, and $\gamma$ is the adiabatic index.%

Because of the symmetry inherent to our system (as shown in Fig.~\ref{fig:initial-fields}) we can assume that $\partial_\phi = \partial_z = 0$. \textcolor{black}{Edge effects begin to come into play near the edges of the cylinder but because the fast magnetosonic wave travels fastest perpendicular to the field, we have found no significant error so long as the we measure within the cylinder's $z$ bounds.} Additionally, the plasma starts in a screw pinch equilibrium given by \cite{Freidberg_2014}: %
\begin{equation}
    \partial_r \left(p_0 + \frac{B_{\phi 0}^2 + B_{z0}^2}{2\mu_0}\right) + \frac{B_{\phi 0}^2}{\mu_0 r} = 0.
\end{equation}
Assuming the toroidal field, $B_{\phi}$, is only due to the current-carrying wire at $r=0$ \textcolor{black}{m} and thus unperturbed by the plasma we find that
\begin{align}
    \partial_r p_0 &= -\frac{1}{\mu_0} B_{z0} \partial_r B_{z0} \\
    p_0(r) &= \frac{B_H^2 - B_{z0}(r)^2}{2\mu_0} \label{eq:pressure} \\
    \Rightarrow \rho_0(r) &= \frac{m_i + m_e}{2 T_0 \mu_0} \left(B_H^2 - B_{z0}(r)^2\right) \label{eq:density}.
\end{align}
Here we have used that the ion temperature is negligible compared to the electron temperature such that $p_0 = n_0 T_e = \rho_0 T_e / (m_i + m_e)$.

We can then write out the $r$, $\phi$, and $z$ components of Eq. \ref{eq:vector_linearized_MHD} as
\begin{equation}
\begin{aligned}
    \rho_0 \partial_{tt} \xi_r &= \frac{1}{2 r^2 \mu_0} \left[ r^2 (B_H^2 \gamma - (\gamma - 2) B_{z0}^2) \xi_r'' \right. \\
    &+ r (B_H^2 \gamma - (\gamma - 2) B_{z0} (B_{z0} + 2r B_{z0}' ) ) \xi_r'\\ 
    &\left. + ( -B_H^2 \gamma + (\gamma - 2) B_{z0} (B_{z0} - 2r B_{z0}')) \xi_r \right] \\
    \partial_{tt} \xi_\theta &= 0 \\
    \partial_{tt} \xi_z &= 0
\end{aligned}
\label{eq:simulation_equation}
\end{equation}
where primes are partial derivatives in $r$. This shows that the plasma only moves in the $r$ direction and thus we only need to simulate $\xi_r$.

We initialize $\xi_r = 0$ \textcolor{black}{m} and $\partial_t \xi_r = 0$ \textcolor{black}{m/s} everywhere, since the plasma begins at rest. Our plasma beta is low ($\beta \approx 0.1$) and centered about $r=0$ \textcolor{black}{m} so we assume the distribution is a Gaussian with peak density of $n_0 = 5 \cdot 10^{18}$ \textcolor{black}{m}$\color{black}^{-3}$ (Fig.~\ref{fig:simulation-setup}, a). In the simulation we use a Gaussian \textcolor{black}{but any distribution may be used. The wavefront speed being the fast magnetosonic speed is true for all distributions.}

At the lower boundary, $r = r_0$, we set $\xi(r_0) = 0$ \textcolor{black}{m}. We use $r_0$ which is slightly greater than 0 to avoid infinities in the differential equation and set $\xi$ to be a constant as the system is symmetric and thus there can not be any displacement at the center. At the upper boundary, $r = r_1$, the system is controlled by the drive cylinder which sets the loop voltage (via a capacitor bank). This then controls the amount of flux that can escape and is given by
\begin{align}
    V_{\textrm{loop}} &= \dv{\Phi}{t} \approx \oint \vb{v} \cross \vb{B}_0 \cdot \dd \vb{l} = \int_{0}^{2\pi} (\vb{v} \cross \vb{B}_0)_\phi r_1 \dd \phi \\
    \Rightarrow v_r &= \partial_t \xi_r (r_1) = -\frac{V_{\textrm{loop}}}{2 \pi r_1 B_{z0}(r_1)}
\end{align}
Here we have assumed that the magnetic field at the boundary does not significantly change, and is the same linearizing assumption we are using in our simulation equations. The system can be well modeled as a \textcolor{black}{step function with some ramp time of the form}
\begin{equation}
    \partial_t \xi_r(r_1, t) = A \left(1 - e^{-t/\tau}\right) 
\end{equation}
where $A$ is the velocity amplitude calculated using the initial voltage of the capacitors ($\sim 1 \cdot 10^{\color{black} 5}$ m/s), and  $\tau$ is the rise time due to transmission line effects for the increase ($\sim 2 \cdot 10^{-7}$ s).

Fig.~\ref{fig:simulation-setup} shows typical simulation results. Panels a) and b) show the initial plasma distribution and corresponding wave speed of a fast magnetosonic wave \textcolor{black}{propagating perpendicular to the magnetic field} \textcolor{black}{with speed given by}
\begin{equation}
    \color{black} v_{ms} = \sqrt{v_A^2 + c_s^2}
    \label{eq:magnetosonic}
\end{equation}
\textcolor{black}{where $v_A$ is the Alfv\'en velocity and $c_s$ is the ion sound speed.} Panel d) shows a streamline traveling at $v_{ms}$, which travels with the wavefront caused by the initial change in the boundary velocity (panel c)). Finally, panel e) shows the expected $\partial_t B_z$ signal that matches well with the measured signal from Fig.~\ref{fig:experimental-evolution}, panel a).

\section{Data Evaluation}\label{sec:data_evaluation}

Given that the simulation and experiment qualitatively match, we \textcolor{black}{now describe in more details the analysis of the measured $\partial_t B_z$ data.} \textcolor{black}{We measure the wave speed by setting a threshold on $\partial_t B_z$ and calculating the crossing time for each probe. Then, using a central finite-difference derivative, we calculate the wavefront speed. We sample a uniform distribution of threshold values to account for noise and non-linear wavefront shape. From this, we find a mean and standard deviation of the measured velocity, $v_{\textcolor{black}{\text{meas}}}$, for each probe with an example threshold result shown in Fig.~\ref{fig:wave-vs-probe}. We then combine this with the error due to probe position, timing accuracy, temperature estimation, and background field strength to achieve the final error bars. The shock and reconnection layer are approximate locations where the field strength becomes stronger and the background field reverses directions respectively.}

Once the speed of the wave is known, the radial density profile is \textcolor{black}{calculated using}
\begin{equation}
\label{eq:magneto}
    \color{black} n_0 = \frac{B_H^2}{2\mu_0} \frac{1}{\frac{1}{2}m_i v_{\text{meas}}^2 + (1 - \gamma / 2) T_e}
\end{equation}
\textcolor{black}{where we have solved for $n$ after combining equations \ref{eq:density} and \ref{eq:magnetosonic}.} The magnetosonic speed depends on the magnetic field magnitude, and the electron temperature and density. The electron temperature is set to 4 eV for \textcolor{black}{density calculations} which is \textcolor{black}{approximately what is} measured by Langmuir probes.%

\textcolor{black}{Since the temperature changes as a function of radius by up to a factor of two, the assumption of constant temperature introduces error into the the derived $n_0$. It can be seen that $n_0 \propto 1 - (1 - \gamma/2) \beta$ which for our low beta system introduces errors of up to 2\% due to temperature differences. Setting $T_e = 0$ eV results in the assumption of the wave moving at the Alfv\'{e}n speed which is accurate so long as the actual plasma beta is small}. \textcolor{black}{Again, this approximation is reasonable for our application, but using Eq.~\ref{eq:magneto} in place of Eq.~\ref{eq:n0} is of course straight forward.}

\section{Procedure Validation}\label{sec:procedure_validation}

\begin{figure}
    \includegraphics{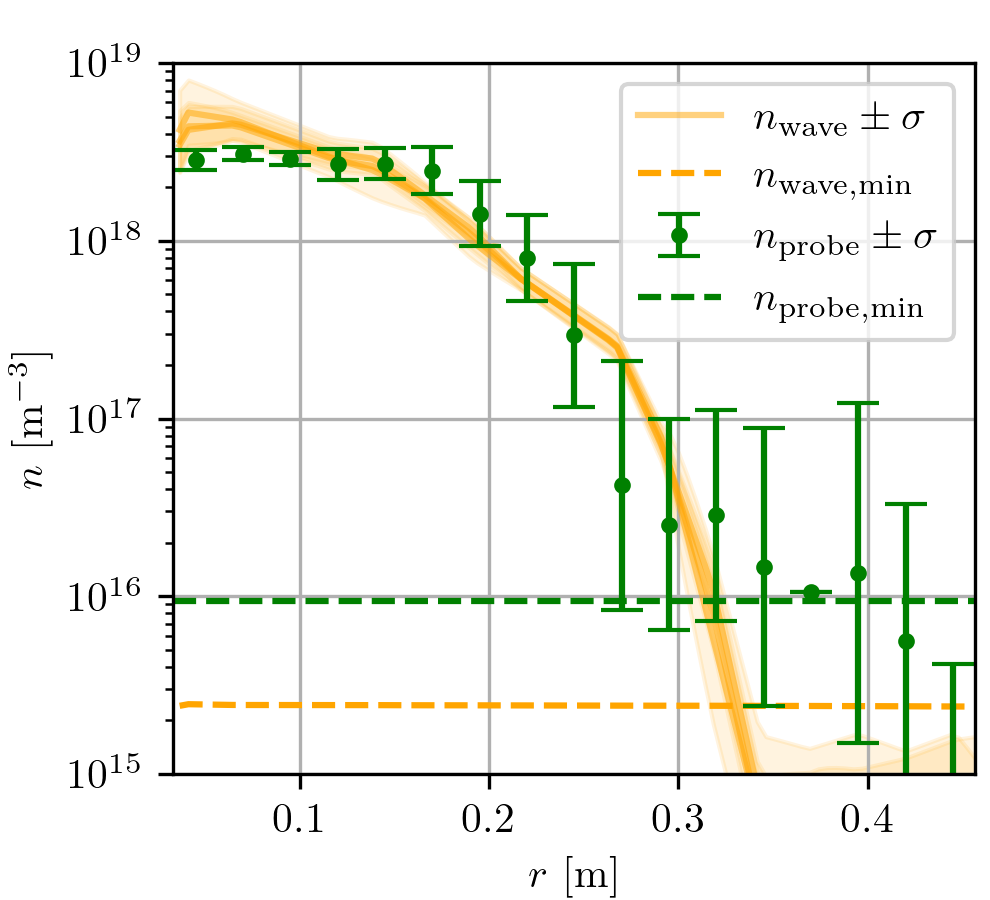}
    \caption{Results from density measurements using the Langmuir probe (green dots) and the wave which have been interpolated in space (orange lines). Each green point is a single shot whereas each orange line is constructed per shot and four shots are overlaid \textcolor{black}{with one standard deviation error bars}. Density measurements by the pressure wave are very reproducible between shots. The green dashed line shows the minimum measurable density for the Langmuir probe which is set by the Debye length vs. the tip separation. The orange dashed line sets the minimum measurable density using the wave which is set by the probe separation, sampling frequency, and magnetosonic speed.}
    \label{fig:wave-vs-probe}
\end{figure}

\textcolor{black}{To verify the accuracy of the inferred plasma density profiles}, we have used a Langmuir probe\cite{olson_2020} to calculate the density at various radial positions, as seen in Fig.~\ref{fig:wave-vs-probe}.
Although some imperfections in the matching exist, we can see that the density between the two methods agrees very well at the center of the device. Meanwhile, at high radii where the density is low, the speed of the wavefront is such that multiple probes see the front within the same sampling period, which thus sets the minimum measurable density using this technique, $n_{\textrm{wave,min}}$. Our Langmuir probe also becomes unreliable at such low densities. This sets the probe's minimum measurable density, $n_{\textrm{probe,min}}$, as shown in Fig.~\ref{fig:wave-vs-probe}. \textcolor{black}{Thus, this method is only validated for sufficiently high densities.}

\begin{figure}
    \includegraphics{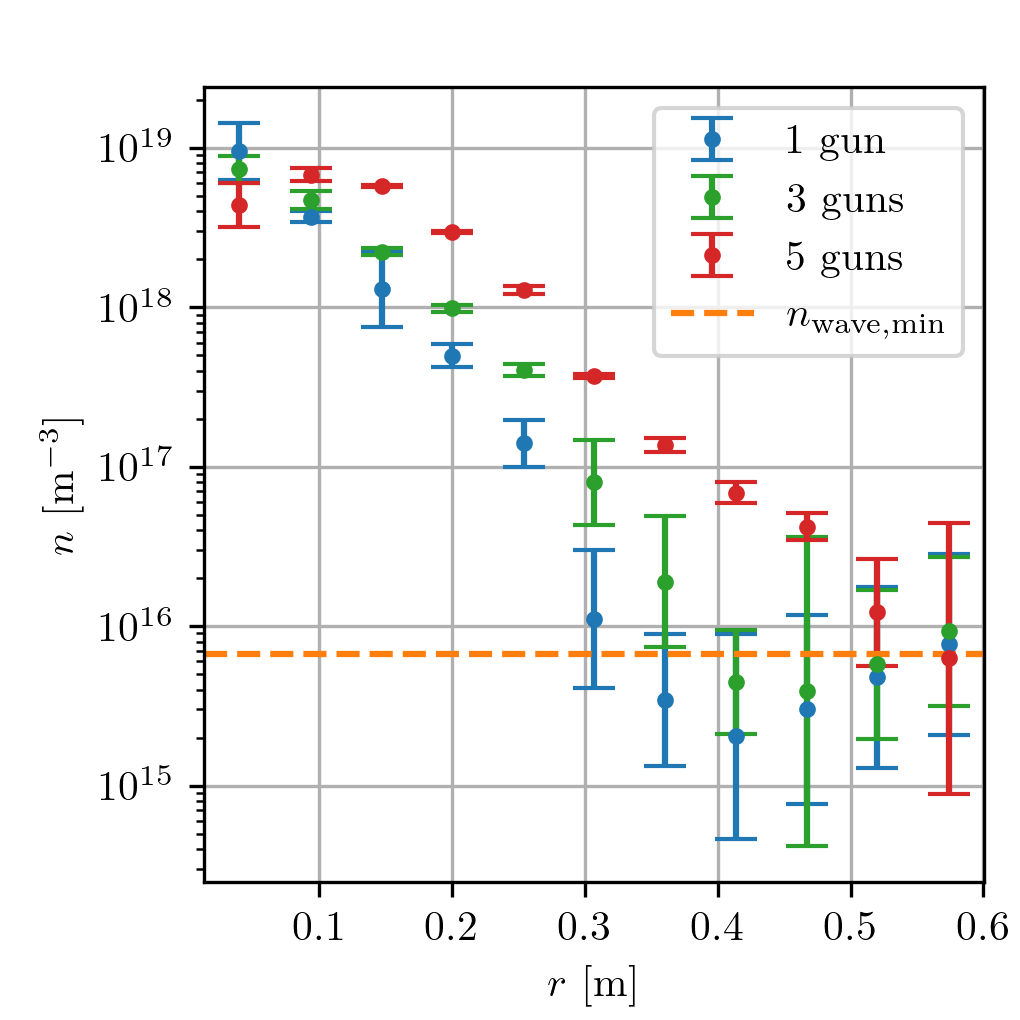}
    \caption{By taking shots with different numbers of plasma guns we can achieve a variety of equilibrium plasma profiles. From this data measured with the pressure wave analysis, we see that with more guns, the density is higher in total and has a larger spread although the peak density stays similar. \textcolor{black}{The error bars are one standard deviation.}}
    \label{fig:varying-gun}
\end{figure}

As another check, we have measured the density profile for various numbers of plasma guns feeding plasma into the BRB. As expected physically, injecting more plasma increases the total density, as seen in Fig.~\ref{fig:varying-gun}. %
A major advantage of our new method for characterizing the initial density is the ability to measure the entire density profile for each shot. This is in contrast to our previous approach where the radial scan was implemented by moving the Langmuir probe between shots. Furthermore, the magnetic diagnostics, by which the magnetic field dynamics are recorded, are designed and optimized for recording the magnetic field evolution during the full magnetic reconnection experiments.

\section{Conclusions}\label{sec:conclusions}

In the present paper, we have developed a new approach for characterizing the profile of the plasma density in the BRB. The plasma is created by plasma guns and is peaked at the center of the device. As a part of the magnetic drive applied in magnetic reconnection experiments, a large amplitude magnetosonic wavefront propagates radially inwards
in a near perfectly cylindrical configuration. For each plasma discharge, the propagation speed of the wave is recorded by {\sl in situ} magnetic diagnostics along radial chords. This provides a highly efficient method for characterizing the initial number density profile at the time before the profile is modified by the subsequent dynamics of magnetic reconnection.

Numerical simulations reproduce with accuracy the experimental data and also agree with the intuitive behavior where the information of the plasma expansion is carried by the magnetosonic wave. In addition, the inferred plasma density profiles are in agreement with similar profiles recorded by Langmuir probes. 
The method is general in nature, and could possibly also be applied in other cylindrically symmetric experiments including LAPD\cite{gekelman_2016} or MRX\cite{Yamada_2006} \textcolor{black}{so long as coils are used to pulse an axial field.}. The only knowledge needed is the direction and speed of the wavefront traveling at a fast magnetosonic speed.

\section*{Acknowledgments}
We acknowledge DOE fund Grant No. DE-SC0020989 and NASA fund Grant No. 80NSSC22K0556 for the support of the TREX experiment. In addition, the experimental work was supported through the WiPPL Collaborative User Facility under DOE fund Grant No. DE-SC0018266.

\section*{Author Declarations}

\subsection*{Conflict of Interest}
The authors have no conflicts of interest to disclose.
\subsection*{Author Contributions}
\textbf{Cameron Kuchta:} %
Conceptualization (equal); Data Curation (equal); Formal Analysis (lead); Investigation (equal); Methodology (equal); Software (equal); Supervision (supporting); Visualization (lead); Writing -- original draft (equal); Writing -- review \& editing (equal).
\textbf{Jan Egedal:}
Conceptualization (equal); Data Curation (equal); Formal Analysis (supporting); Funding Acquisition (equal); Investigation (equal); Methodology (equal); Project Administration (equal); Supervision (lead); Writing -- original draft (equal); Writing -- review \& editing (equal).
\textbf{Abhishek Mhatre:}
Data Curation (equal); Formal Analysis (supporting); Investigation (supporting); Software (equal); Writing -- original draft (equal).
\textbf{Paul Gradney:}
Conceptualization (supporting); Data Curation (supporting); Investigation (supporting); Writing -- review \& editing (supporting).
\textbf{Joseph Olson:}
Conceptualization (supporting); Data Curation (supporting); Investigation (supporting); Project Administration (equal); Supervision (equal); Writing -- review \& editing (supporting).
\textbf{Xinyu Yu:}
Data Curation (supporting).
\textbf{Cary Forest:}
Conceptualization (supporting); Funding Acquisition (equal); Project Administration (equal); Resources (lead).
\section*{Data Availability}
The data that support the findings of this study are available from the corresponding author upon reasonable request.
\section*{References}
\bibliography{references}

\end{document}